# Raman spectroscopic detection of rapid, reversible, early-stage inflammatory cytokine-induced apoptosis of adult hippocampal progenitors/stem cells

Uma Ladiwala[1,*], Aseefhali Bankapur[2], Bhushan Thakur[1],

Chidangil Santhosh[2], and Deepak Mathur[2,3]

[1] UM-DAE Centre for Excellence in Basic Sciences, Kalina Campus, Mumbai 400 098, India

[2] Centre for Atomic and Molecular Physics, Manipal University, Manipal 576 104, India

[3] Tata Institute of Fundamental Research, 1 Homi Bhabha Road, Mumbai 400 005, India

[*]*e-mail:* brainwave@cbs.ac.in

**Abstract**

The role of neuro-inflammation in diverse, acute and chronic brain pathologies is being increasingly recognized. Neuro-inflammation is accompanied by increased levels of both pro- and anti-inflammatory cytokines; these have deleterious as well as protective/reparative effects. Inflammation has varying effects on neurogenesis and is a subject of intense contemporary interest. We show that TNF-α and IFN-γ, used concomitantly, cause apoptosis of adult rat hippocampal progenitor/stem cells *in vitro* as detected by the TUNEL and MTT assays on time scales of several hours. We have coupled Raman spectroscopy to an optical trap to probe early changes of apoptosis in single, live neural stem cells that have been treated with pro-inflammatory cytokines, TNF-α and IFN-γ. Changes caused by inflammation-induced denaturation of DNA are observed in the Raman spectra that correspond to very early stages of apoptosis, occurring on very fast time scales: as short as 10 minutes. Addition of the anti-inflammatory cytokine IL-10 either 10-30 min before or 10-30 min after treatment with TNF-α and IFN-γ reverses the changes substantially. Our findings imply that inflammation can induce very rapid changes leading to cell death but that these are reversible, in the early stages at least.

**Key Words:** neural stem cell, Raman spectroscopy, cytokine, apoptosis

# 1. Introduction

Brain inflammation is known to be a complex cascade of cellular and molecular responses to stress, injury, infection or neurodegeneration of the central nervous system (CNS). The inflammatory response, which attempts to get rid of, or contain, the stressor and return the CNS to its normal state, occurs within minutes and involves both cells as well as an array of immune mediators like the cytokines.

A commonly observed outcome of neuro-inflammation is apoptosis, that is, cell death of neural cells, which is characterized by DNA condensation and fragmentation, nuclear shrinkage, chromatin and cytoplasmic condensation and disintegration (Kerr et al, 1972, Wyllie et al, 1980). It appears that cells in the hippocampus, one of the regions of ongoing neurogenesis, are particularly vulnerable to both acute and chronic inflammation as occur in traumatic brain injury (TBI) and neurodegenerative disorders. The cognitive deficit with memory impairment and emotional disturbances that is often a sequel to brain inflammation, is due both to neuronal apoptosis as well as damage to neural stem cells (NSCs) with disrupted neurogenesis in the hippocampus (Krawohl & Kaiser 2004 a,b; Peng et al, 2008a; Waldau & Shetty, 2008). Acute inflammation is accompanied by increased levels of pro-inflammatory cytokines such as TNF-α, IL-1β, IL-6 and IFN-γ (Csuka et al, 1999; (Morganti-Kossman et al 2001,Goodman et al, 2008; Frugier et al, 2010; Dalgard et al, 2012; Christie and Turnley, 2013) which have varied effects on neurogenesis (Ben-Hur et al, 2003;Wong et al, 2004; Cacci et al, 2005; Sheng et al, 2005; Iosif et al, 2006; Tzong-Shine et al, 2008). Acute insult also raises levels of anti-inflammatory cytokines like IL-10 (Csuka et al, 1999; Hensler et al, 2000; Dziurdzik et al, 2004; Shiozaki et al, 2005; Kamm K et al, 2006; Kirchoff et al, 2008).

The injured brain attempts repair by producing new neurons. Injury is also amenable to rescue by endogenous or exogenous therapeutic interventions such as cell transplantation. For such interventions to succeed, it is crucial to identify very early time points at which changes of cell death or damage occur post-injury. It has hitherto been observed, using standard biological assays, that such changes occur on time scales of several hours to days (Conti et al, 1998; Holmin & Hojeberg 2004; Sheng et al, 2005; Cacci et al, 2005). In experiments whose results we report in the following we have made use of microRaman spectroscopy (RS) coupled to an optical trap to detect and quantify early changes of apoptosis in single, live neural stem cells that have been treated with pro-inflammatory cytokines, TNF-α and IFN-γ. Our methodology is such that we are able to detect, for the first time, the effect of pro- and anti-inflammatory cytokines on neural stem cells on time scales as short as 10 minutes. We observe that addition of the anti-inflammatory cytokine IL-10 either 10 min before or 10 min after treatment with TNF-α and IFN-γ reverses the changes substantially. Our findings imply that inflammation can induce very rapid changes leading to cell death, but that these are reversible at the early time points tested.

## 2. Materials and Methods
### 2.1 Single cell spectroscopy with Raman tweezers

Raman spectroscopy measurements were carried out using a home built single-beam Raman Tweezers set-up whose details have been published elsewhere (Bankapur et al, 2010; Zachariah et al, 2010). A schematic representation is shown in Fig. 1A along with a typical Raman spectrum of the NSC (Fig. 1B) recorded by keeping a slit width of 100 μm. The Raman spectrum was measured using a single laser beam (785 nm wavelength) to optically trap, and excite, a single live cell for Raman scattering. The spatial intensity distribution within the laser beam was expanded from nearly 1.5 mm in diameter (the size of the beam

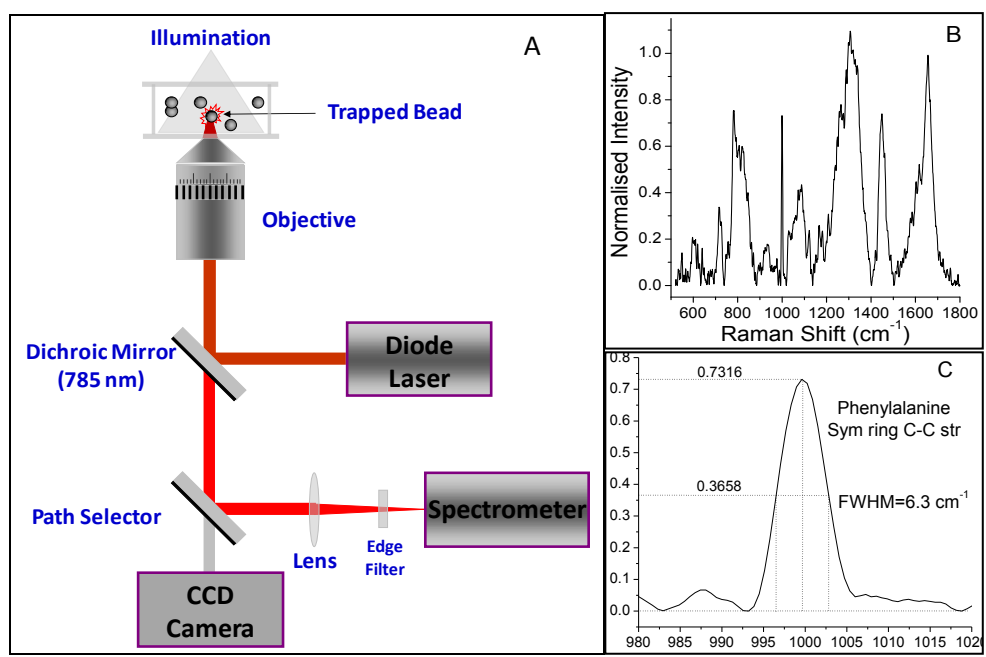

Figure 1. A: Schematic representation of the single beam Raman tweezers set-up. B: Raman spectrum of a single, live cell. C: Determination of spectrometer resolution with respect to phenylalanine peak at 999 cm$^{-1}$ in the above spectrum.

emanating from our diode laser) to 10 mm using a Galilean type beam expander comprising a plano-concave lens (f=5 cm) and a biconvex lens (f=30 cm). A dichroic mirror with high reflectivity at 785 nm and high transmission above 785 nm directed the beam vertically into the microscope objective of an inverted microscope (Nikon Eclipse Ti-U, Japan) such that the expanded beam overfilled the objective (Fig. 1A). The objective had a large numerical aperture (1.3 NA) and 100x magnification (Nikon, Japan) which enabled the laser beam to be focused to a near-diffraction limited spot, thus facilitating the formation of a high intensity gradient that is crucial for optical trapping of single cells (Ashkin, 1970)

Backscattered light was collected by the same microscope objective and was directed to the output port of the microscope by an optical path selector. The weak Raman signal was separated from the accompanying strong Rayleigh signal by means of an optical edge filter (Semrock, USA). The filtered Raman signal was focused onto the slit of an f/4.1 spectrograph (iHR 320, Horiba Jobin Yvon, Japan) equipped with a holographic grating (1200 grooves/mm) which laterally dispersed the signal. A concave mirror (f=32 cm) focused the laterally separated spectral lines (the Raman lines constituting the spectrum) on to a liquid nitrogen cooled CCD detector (Symphony, Horiba Jobin Yvon, Japan). We determined the resolution of our spectrometer with respect to the phenylalanine peak at 999 cm$^{-1}$ of the cell spectrum (Fig. 1C) to be 6.3 cm$^{-1}$.

Our measurement procedure in the present series of experiments was as follows. Raman spectra were recorded by us using NSCs before and after treatment with the inflammatory cytokines TNF-α and IFN-γ, with and without the anti-inflammatory cytokine IL-10. Adult rat hippocampal NSCs/progenitors were suspended in phosphate buffer saline (PBS). 200 µl of this suspension was placed in a custom-made sample cell and control Raman spectra were acquired by trapping a single cell from the suspension. After acquisition of Raman data, 20 ng/ml of TNF-α and IFN-γ were added to this 200 µl cell suspension. Care was taken to retain the cell from which the control Raman spectrum was recorded within the trapping volume while adding the cytokines. The trapped cell was left for 10 minutes under the effect of the added cytokines following which three sets of Raman spectra (10 minutes acquisition time in each case) were recorded consecutively from the same cell. The experiment was repeated by adding 20 ng/ml of IL-10, either 10 minute pre- or post-treatment with TNF-α + IFN-γ (20 ng/ml each). The three experimental conditions utilized in our series of measurements are as summarized in Table I.

To summarize, we recorded (a) four spectra each from a single, live, trapped cell in every condition, (b) one spectrum before adding the cytokines (baseline control), and (c) three successive spectra 10 minutes after addition of the cytokines. Control experiments were also performed by recording three successive Raman spectra every 10 minutes from a cell without adding any of the cytokines.

Table I. Summary of the measurement procedure followed in the present experiments.

| Condition | Cytokines added after recording baseline control spectrum | | Raman Spectra Recorded |
|---|---|---|---|
| | 0$^{th}$ Minute | 10$^{th}$ Minute | |
| I | TNF-α+IFN-γ | --- | Control, Cycle11, Cycle12, Cycle13 |
| II | IL-10 | TNF-α+IFN-γ | Control, Cycle21, Cycle22, Cycle23 |
| III | TNF-α+IFN-γ | IL-10 | Control, Cycle31, Cycle32, Cycle33 |

**2.2 Isolation of hippocampal progenitors/stem cells, TUNEL and MTT assays**

Hippocampal progenitors/stem cells were isolated from Wistar rats (wt. 200-250 gm) according to a modification of an earlier described protocol (Palmer et al, 1995). Briefly, hippocampi were dissected from brains of the rats and neural progenitor cells were isolated using a Percoll density gradient centrifugation. To obtain fairly pure populations of neural progenitors, the isolated cells were initially plated on poly-L-ornithine (20 µg/ml) and laminin (10 µg/ml) - coated T-25 flasks in DMEM/F-12 medium supplemented with B-27 and 40ng/ml FGF-2 at 37˚C in a humidified atmosphere with 5% $CO_2$. After about 3 weeks, dense colonies of proliferating progenitor-like cells were manually stripped, mechanically dissociated and plated in poly-L-ornithine- coated T -25 flasks (Nunc, USA) in Neurobasal medium supplemented with B-27 and 40 ng/ml FGF-2. NSCs were characterized by immunocytochemical detection of marker expression as shown in Supplementary Fig 1.

To ascertain the effects of cytokines on survival/cell death of adult rat hippocampal progenitors in vitro, the TUNEL (Terminal deoxynuceotidyl transferase dUTP nick-end labelling) assay was performed using the TACS 2 TdT-Fluor in situ apoptosis detection kit (Trevigen, USA) on NSCs plated in Lab-Tek (USA) chamber slides treated in triplicate wells with the following concentrations of cytokines: IL-1β (20 ng/ml); TNF-α (20 ng/ml); IFN-γ (20 ng/ml); IL-6 (50 ng/ml), IL-10 (20 ng/ml), IL-4 (20 ng/ml), TGF-β (10 ng/ml) and combinations thereof, for a period of 48 or 72 hours. For the MTT (3-4, 5-dimethylthiazol-2-yl-2,5-diphenyltetrazolium bromide) reduction assay for cell viability/survival, NSCs were plated in 96-well tissue culture plates and treated with cytokines, with 6 wells per condition. At the end of each experiment medium from each chamber- slide well was removed and 100 µl MTT reagent (0.5 mg/ml in DMEM) added in the dark and plates incubated overnight in the $CO_2$ incubator at 37C. The reaction was then stopped by addition of 100 µl/ well of 10% SDS- 0.01 N HCl and kept in the dark for 6 hours on a slow shaker. The supernatant from each well was analyzed spectrophotometrically using a TECAN ELISA plate reader (Switzerland) at 570 nm wavelength. Readings from the different conditions were expressed as percent viability with respect to control.

## 3. Results and discussion

A typical Raman spectrum of an NSC is shown in Fig. 2 with Raman frequencies assigned to the multifarious peaks that are clearly resolved. We utilized available literature to carry out the frequency assignment (Bankapur et al, 2010; Parker, 1983). Table II lists the assigned frequencies for the peaks in our spectra; the tabulated data provide an overview of the spectral features that characterize the stem cells used in our experiments. Although a detailed analysis of the individual Raman features is not within the scope of the present report, we note that the Raman spectrum is dominated by lines that correspond to signatures expected from proteins, lipids, and nucleic acids - the major building blocks of cells. For the purposes of the present set of experiments, we focus on those parts of the spectrum that show differences before and after addition to the cell suspension of the inflammatory cytokines, TNF-α and IFN-γ, and anti-inflammatory cytokines IL-10. These differences are clearly seen in Fig. 3 where we overlay spectra measured prior to and post-treatment.

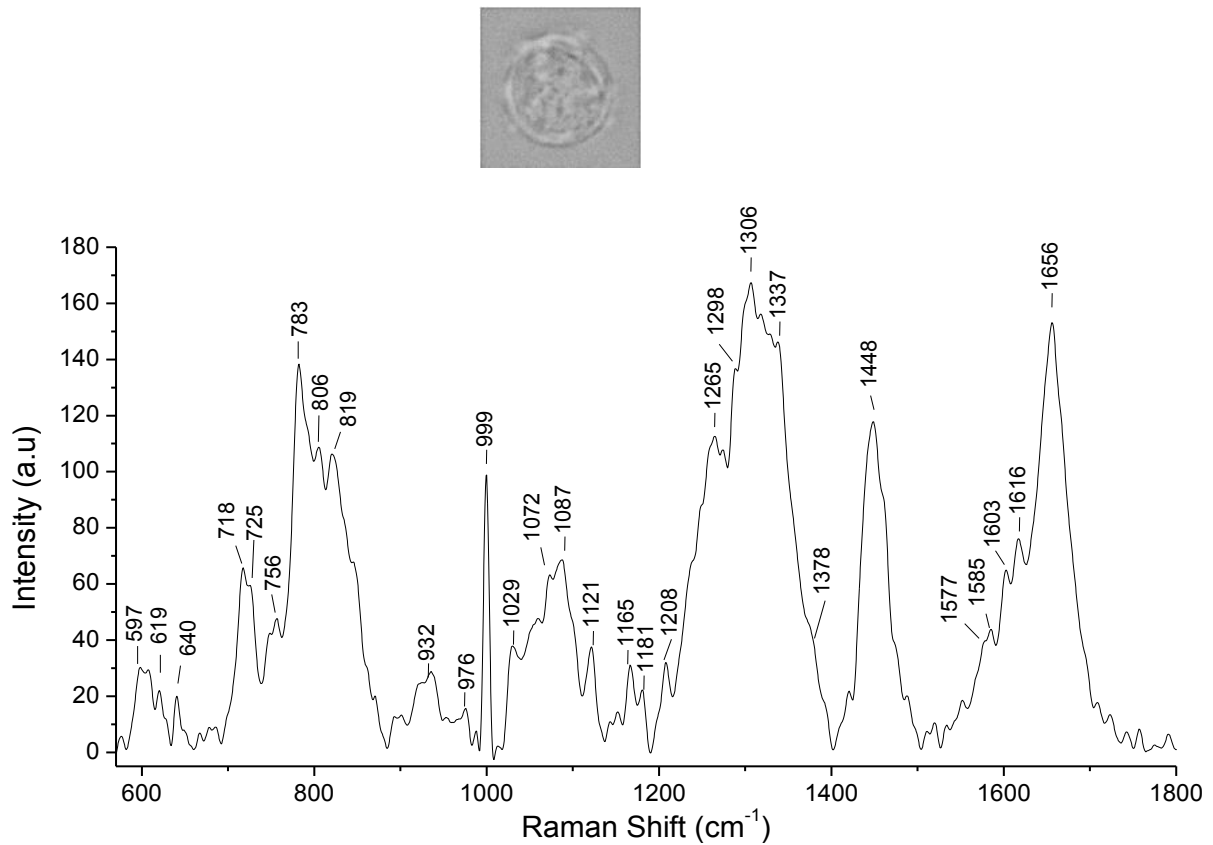

Figure 2. Typical Raman spectrum of an optically trapped neural stem cell (inset) recorded with 25 mW of laser power and 60 sec of acquisition time while giving 5 average accumulations.

The overlaid Raman spectra (Fig. 3 A-C) include the control spectrum (prior to treatment) and spectra measured post-treatment (cycles 11, 12, and 13) with cytokines TNF-α and IFN-γ. A comparison reveals an increase in the intensity of the Raman band located in the region 775-875 cm$^{-1}$ which corresponds to nucleic acids and lipids. The increase in intensities of lines corresponding to nucleic acids is most likely due to the inflammation induced denaturation of DNA initiated by the cytokines. We have monitored the modification of this Raman signature under the influence of IL-10, an effective anti-inflammatory agent and typical results are shown in Fig.3D.

TABLE II. Assignment of peaks observed in the Raman spectrum shown in Figure 2.

| Sl.No | Frequency (cm$^{-1}$) | Assignments |
|---|---|---|
| 1 | 597 | p: Amide VI (membrane) |
| 2 | 619 | Phe :C-C twist |
| 3 | 640 | p: C-S str, Tyr: C-C twist |
| 4 | 718 | Phospholipid:Choline: C-N symstr |
| 5 | 725 | A |
| 6 | 756 | T,Trp |
| 7 | 783 | C,T,U, DNA or Lip:O-P-O diester sym str |
| 8 | 806 | DNA:bk:O-P-O sym str |
| 9 | 819 | Phospholipid: O-P-O diesterantisymstr |
| 10 | 932 | p: skeletal vibration |
| 11 | 976 | Deoxyribose |
| 12 | 999 | Phe: C-C skeletal vibration |
| 13 | 1029 | Phe, p: C-N str |
| 14 | 1072 | Phospholipid: C-O str |
| 15 | 1087 | DNA:bk:O-P-O sym str, p:C-N str, Phophoryl Cholin: PO$_3^{-2}$ anti-sym str |
| 16 | 1121 | p: C-N str |
| 17 | 1165 | Collagen: NH$_3^+$ |
| 18 | 1181 | Tyr, NA base external C-N Str, Phe, p:C-H bend |
| 19 | 1208 | Tyr, Phe, A, T, p: Amide III |
| 20 | 1265 | p: Amide III, C, A |
| 21 | 1298 | p: Amide III |
| 22 | 1306 | A, p: Amide III,Phospholipid:CH$_2$ |
| 23 | 1337 | A, G, p: C-H def |
| 24 | 1378 | T,A,G |
| 25 | 1448 | p:C-H$_2$ def |
| 26 | 1577 | A, G |
| 27 | 1585 | A, G |
| 28 | 1603 | Phe, Tyr, p: C=C |
| 29 | 1616 | Tyr, Trp, p: C=C |
| 30 | 1656 | p: Amide I |

**Abbreviations**: def: deformation, bk: vibration of DNA backbone, Str: stretching, U,C,T,A,G:ring breathing modes of the DNA/RNA bases, Tyr: Tyrosine, Trp: Tryptophan, Phe: Phenylalanine, p: Protein, Lip: Lipid Rib: ribose, NA: nucleic acids.

Cell treatment with IL-10 prior to addition of TNF-α and IFN-γ appears to reverse the Raman band intensity in the 775-875 cm$^{-1}$ region: the intensity of the band decreases with respect to that measured in the control spectrum (Fig. 3D). The extent of the decrease is significant, even on time scales as short as 10 minutes (>20% decrease), with nearly a 40% decrease obtained after 30 minutes.

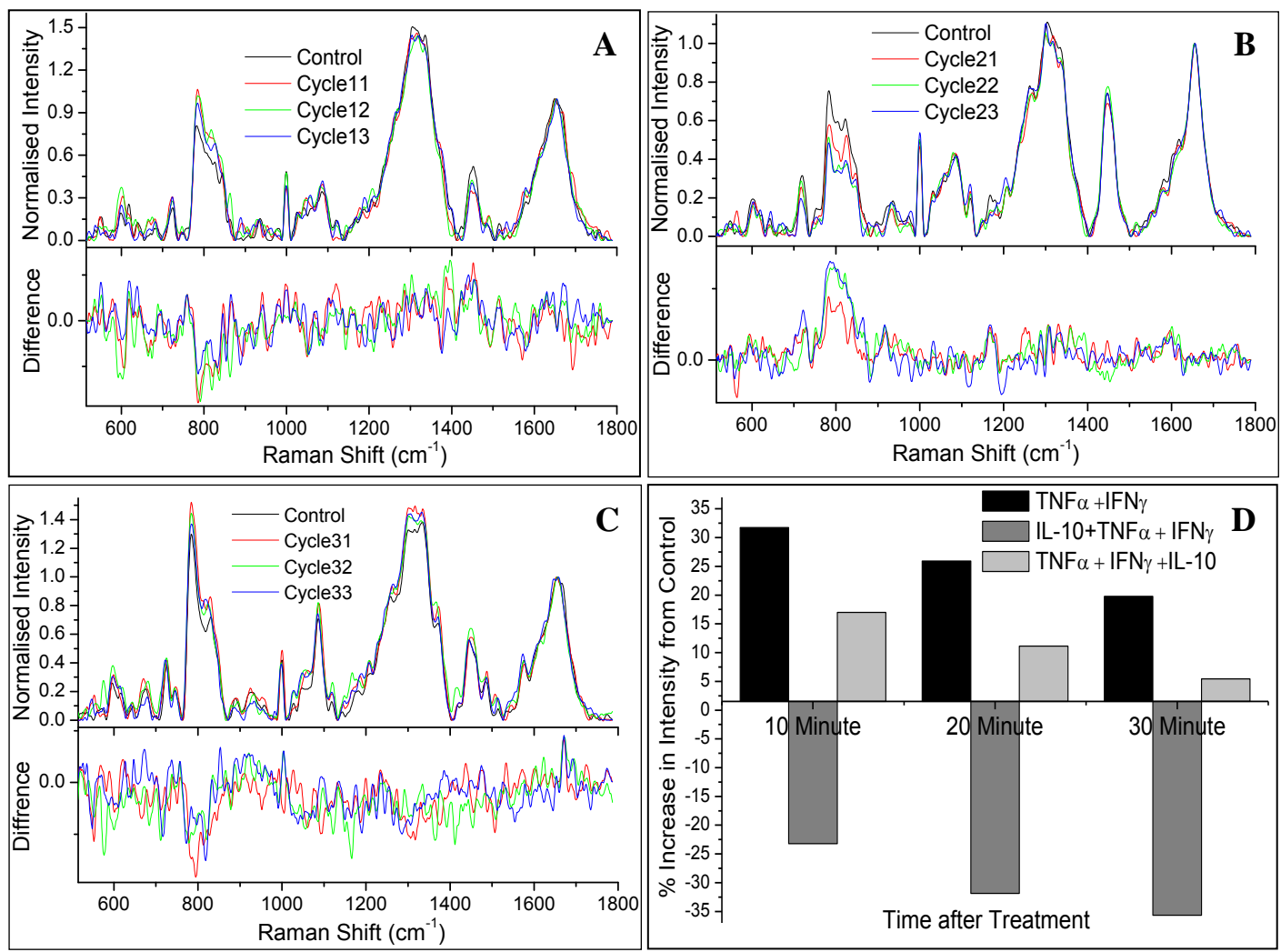

Figure 3. Overlaid Raman spectra of NSCs before (control) and after treatment with cytokines: A: (TNF-α + IFN-γ); B: IL-10 + (TNF-α + IFN-γ); C: (TNF-α + IFN-γ) + IL-10; D: Intensity variations of the 783 cm$^{-1}$ peak corresponding to DNA in the above three conditions. The lower panel spectra in A, B, and C are difference spectra, that is, control spectra – cytokine treatment spectra.

We also made measurements after IL-10 treatment of the cell that was already under the influence of TNF-α and IFN-γ. Here also we observed an increase in the intensity of the Raman line corresponding to the nucleic acid, albeit the increase being somewhat less (Fig. 3): 15% increase after only 10 minutes, reducing to an increase of ~5% after 30 minutes.

There is an acute paucity of earlier work with which we can compare our present observations. It is known that a common outcome of neuro-inflammation consequent on, for instance, traumatic brain injury is apoptosis of neural cells. This is usually characterized by structural changes in DNA, nuclear shrinkage, chromatin and cytoplasmic condensation and disintegration. A Raman spectrum of chemically-fixed lung fibroblast cells under stress due to exposure to glyoxal has provided evidence of a decreased intensity of DNA bands in the 786 and 1003 cm$^{-1}$ region attributed to early changes of apoptosis; it has been hypothesized that shrinkage and increased condensation might lead to more base-base interaction and consequently, decreased DNA intensities (Krafft et al, 2006). Decreased DNA content, as measured in Raman spectroscopy of apoptotic leukemia cells, was deduced from reduced intensity of peaks around 794 cm$^{-1}$ in a recent study by Ong et al (2012); these were taken to indicate fragmented DNA with formation of apoptotic bodies. Our spectroscopic findings of an increase in the intensity of the Raman band located in the region 775-875 cm$^{-1}$, corresponding to nucleic acids and lipids, are consonant with those reported by Notingher et al (2010) who studied human breast cancer cells and showed a ~1.5 fold increased intensity of Raman bands associated with DNA in all cells exposed to 6 hours of etoposide; this increase in intensity was assigned to nuclear condensation seen in the early stages of apoptosis. In our study these changes were seen in IFN-γ and TNF-α - treated NSCs at as early a stage as 10-30 minutes; interestingly, these changes were seen by us to be partially reversed upon addition of IL-10 either 30 minutes before or 30 minutes after treatment with IFN-γ and TNF-α.

We used the TUNEL assay to quantitatively screen for cell death of NSCs exposed to various pro- and anti-inflammatory cytokines *in vitro*. From our results (Fig. 4) it is evident that concomitant addition of IFN-γ and TNF-α caused a significant amount of NSC apoptosis - greater than that seen with either cytokine added singly or with any of the other pro-inflammatory cytokines that we tested. When time-course experiments were performed with IFN-γ and TNF-α, significant apoptosis was detectable only at 18-24 hours by the TUNEL and MTT assays, respectively (Fig 4 C,D), in concordance with other studies (Ben-Hur et al, 2003; Cacci et al, 2005). When the anti-inflammatory cytokine IL-10 was added either 30 min prior to or 30 min after IFN-γ and TNF-α, we observed partial reduction of this outcome as well as improved viability of the NSCs (Fig 4E). This was not evident when IL-10 was added later, at 1 hour, 2 hours, or 3 hours post-treatment (Fig 4E). Until recently apoptosis was considered to be an irreversible process, especially once the critical checkpoints such as mitochondrial fragmentation and DNA damage occurred. In contrast, a few recent studies have provided indications that apoptosis may be reversible either in the early stages (Geske et al, 2001) or even in late stages (Tang et al, 2012).

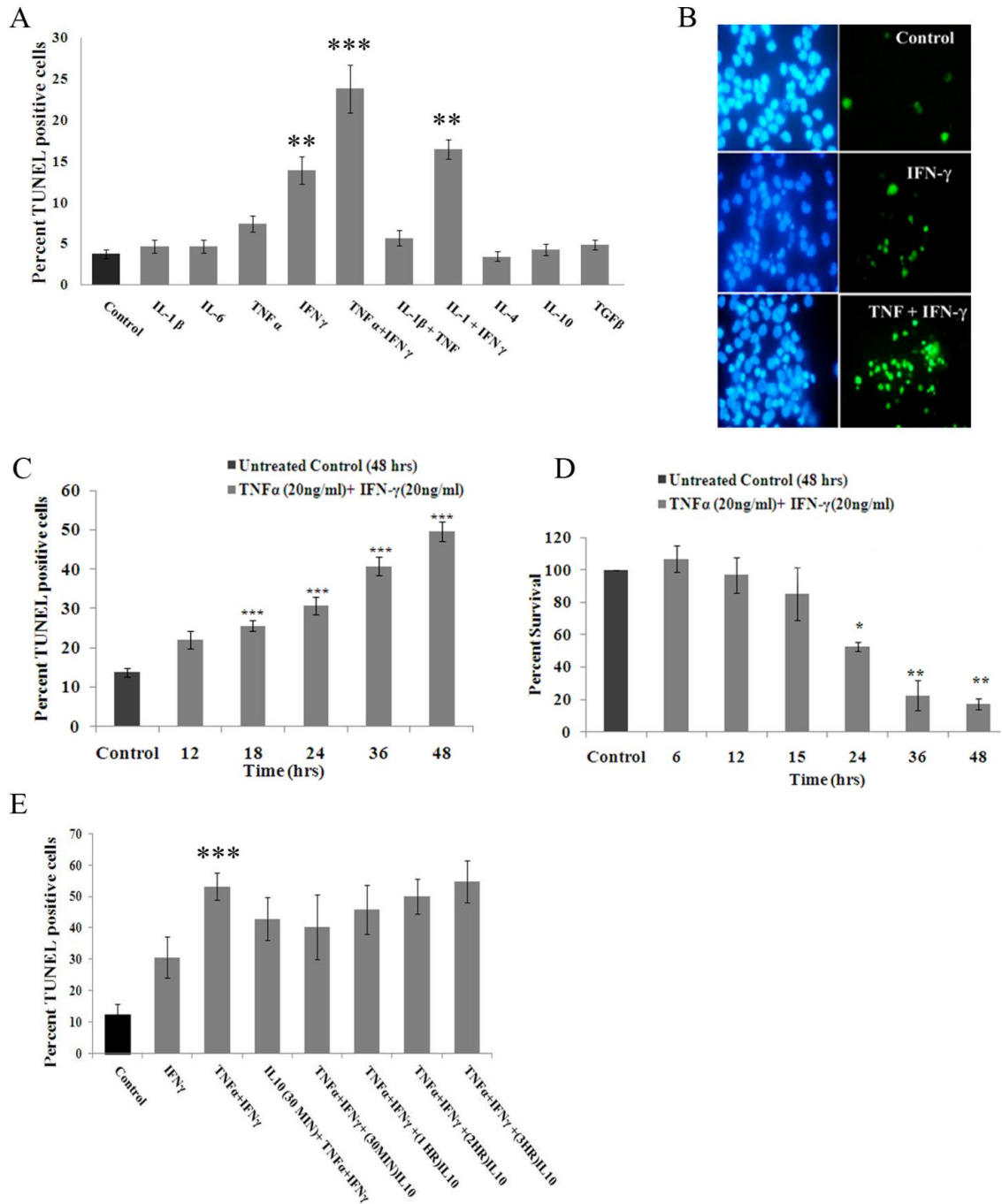

Figure 4. A: Pro-inflammatory cytokines IFN-γ and TNF-α used concomitantly cause significant apoptosis (p< 0.001) of NSCs that can be partly reversed by addition of the anti-inflammatory cytokine IL-10 either 30 min prior or 30 min after treatment with IFN-γ and TNF-α *in vitro*. B: Quantitation of the pro- apoptotic effect of various pro-inflammatory and anti-inflammatory cytokines on NSCs by the TUNEL assay over 72 hours *in vitro*. C,D: Morphology of untreated (control), IFN-γ-treated and IFN-γ + TNF-α - treated TUNEL positive NSCs. E: Viability of IFN-γ and TNF-α- treated NSCs over time by the TUNEL and MTT assays, respectively. Addition of IL-10 to NSCs 30 min prior, and 30 min, 1 hour, 2 hours, and 3 hours after IFN-γ and TNF-α treatment *in vitro.* The pro-apoptotic effect on NSCs was partly reversed only when IL-10 was added 30 min prior (p < 0.05) or 30 min after (p < 0.01).

Neural stem cells have been reported to express receptors for both IFN-γ and TNF-α (TNF-RI, TNF-RII) but not IL-1 or IL-6 (Ben-Hur et al, 2003; Iosif et al, 2006). Ben-Hur et al (2003) reported a decrease in the proliferation of NSCs isolated from newborn rat striata, when both IFN-γ and TNF-α were added together, whereas IFN-γ alone caused increased apoptosis that was partially blocked by the addition of TNF-α. Other studies have reported that TNF-α caused increased apoptosis when added to cells of a neural stem cell line (Cacci et al, 2005) or in fetal neural stem cells through TNF-RI (Sheng et al, 2005), or resulted in cell death of NSCs by necrosis (Wong et al, 2008). IFN-γ inhibited NSC proliferation and differentiated NSCs but did not cause apoptosis in one report (Wong et al, 2004) whereas in another, NSCs and their differentiated progeny were found to express both types of IFN-γ receptors and IFN-γ treatment of NSCs resulted in the development of dysfunctional differentiated progeny (Walter et al, 2011). These apparently conflicting results, including our finding of a pro-apoptotic effect of IFN-γ as well as the additive effect of concomitant IFN-γ and TNF-α, could possibly be an outcome of cells in these studies being isolated from differing species of different ages, from different regions (that is, subventricular zone or subgranular zone of the hippocampus) and whether the experiments were conducted *in vitro* or *in vivo*. It general, it is accepted that suppressing the effects of IFN-γ and TNF-α is beneficial in improving stem cell therapies (Breton and Mao-Draayer, 2011). The anti-inflammatory effects of IL-10 have only recently been studied. IL-10 is also increased in neuro- inflammation (Csuka et al, 1999; Hensler et al, 2000; Dziurdzik et al, 2004; Shiozaki et al, 2005; Kamm K et al, 2006; Kirchoff et al, 2008), and counteracts the deleterious effects of pro-inflammatory cytokines such as IFN-γ and IL-17 (Knoblach & Faden, 1998; Yang et al, 2009). IL-10 increases both neurogenesis and oligodendrogenesis *in vivo* (Yang et al, 2009). More recently, adeno-associated virus-mediated overexpression of neuronal IL-10 ameliorated cognitive dysfunction in a mouse model of Alzheimer's disease (Kiyota et al, 2012) and EAE, the animal model of MS, was suppressed when IL-10-transduced NSCs were injected intravenously (Klose et al, 2013). When IL-10 was administered by different routes in a rodent model of TBI post-injury (Knoblach & Faden, 1998), the outcome was improved only by systemic administration *in vivo* but not when IL-10 was given intra-cerebroventricularly. Our findings show that IL-10 has a direct anti-inflammatory effect in counteracting the pro-apoptotic effects of pro-inflammatory cytokines, IFN-γ and TNF-α, on NSCs *in vitro*. Interleukin-10 has powerful immune-regulatory properties and recombinant IL-10 has been tried for the therapy of established immune diseases with limited success, warranting further investigation (Assadullah et al, 2003).

It is pertinent to note that the primary significance of our findings lies in the recognition that inflammation-induced damage in neural stem cells can occur extremely rapidly and that it can be reversed by early anti-inflammatory measures, such as IL-10 in our study. In acute inflammatory conditions, such as TBI, the inflammatory response begins within minutes - much earlier than was previously thought, and this is accompanied by an increase in levels of many of the pro- and anti-inflammatory cytokines (Kamm et al, 2006; Frugier et al, 2010) thus leaving a narrow window for endogenous or exogenous interventions to prevent significant damage to neural stem/progenitors of the hippocampus. Further investigations, especially in animal models, are required to verify these findings in vivo as well as on the prognosis of clinical symptoms secondary to neuro-inflammation, such as cognitive deficits.

**Figure Legends**

**Figure 1.** A: Schematic representation of the single beam Raman tweezers set-up. B: Raman spectrum of a single, live cell. C: Determination of spectrometer resolution with respect to phenylalanine peak at 999 cm$^{-1}$ in the above spectrum.

**Figure 2.** Typical Raman spectrum of an optically trapped neural stem cell (inset) recorded with 25 mW of laser power and 60 sec of acquisition time while giving 5 average accumulations.

**Figure 3.** Overlaid Raman spectra of NSCs before (control) and after treatment with cytokines: A: (TNF-α + IFN-γ); B: IL-10 + (TNF-α + IFN-γ); C: (TNF-α + IFN-γ) + IL-10; D: Intensity variations of the 783 cm$^{-1}$ peak corresponding to DNA in the above three conditions. The lower panel spectra in A, B, and C are difference spectra, that is, control spectra – cytokine treatment spectra.

**Figure 4.** A: Pro-inflammatory cytokines IFN-γ and TNF-α used concomitantly cause significant apoptosis (p< 0.001) of NSCs that can be partly reversed by addition of the anti-inflammatory cytokine IL-10 either 30 min prior or 30 min after treatment with IFN-γ and TNF-α *in vitro*. B: Quantitation of the pro- apoptotic effect of various pro-inflammatory and anti-inflammatory cytokines on NSCs by the TUNEL assay over 72 hours *in vitro*. C,D: Morphology of untreated (control), IFN-γ-treated and IFN-γ + TNF-α - treated TUNEL positive NSCs. E: Viability of IFN-γ and TNF-α- treated NSCs over time by the TUNEL and MTT assays, respectively. Addition of IL-10 to NSCs 30 min prior, and 30 min, 1 hour, 2 hours, and 3 hours after IFN-γ and TNF-α treatment *in vitro*. The pro-apoptotic effect on NSCs was partly reversed only when IL-10 was added 30 min prior ($p < 0.05$) or 30 min after ($p < 0.01$).